\documentclass{elsart}

\usepackage{amssymb}
\usepackage{graphicx}
\begin{document}

\begin{frontmatter}



\title{The hydration state of HO$^-$(aq)}



\author[LANL]{D. Asthagiri}
\author[LANL]{Lawrence R. Pratt}
\author[LANL]{J. D. Kress}
\address[LANL]{Theoretical Division, Los Alamos National Laboratory, Los
Alamos NM 87545} 
\author[MTHOL]{M. A. Gomez}
\address[MTHOL]{Department of Chemistry, Mount Holyoke College, South Hadley  MA 01075}
\begin{abstract}
The HO$^-$(aq) ion participates in myriad aqueous phase chemical
processes of biological and chemical interest. A molecularly valid
description of  its hydration state, currently poorly understood, is a
natural prerequisite to modeling chemical transformations involving
HO$^-$(aq).  Here it is  shown that the statistical mechanical
quasi-chemical theory of solutions  predicts  that
$\mathrm{HO\cdot[H_2O]_3{}^-}$ is the dominant inner shell coordination
structure for  HO$^-$(aq) under standard conditions.   Experimental
observations and other theoretical calculations are adduced to support
this conclusion.  Hydration free energies of neutral combinations  of
simple cations with HO$^-$(aq) are evaluated and agree well with
experimental values.
\end{abstract}

\end{frontmatter}


\section{Introduction}

In this Letter, we consider the hydration state of HO$^-$(aq) from the
perspective of the quasi-chemical
theory\cite{MartinRL:Hydfiw,HummerG:MulGme,PrattLR:Quatal,Pratt:ES:99,%
HummerG:Moltas,HummerG:Newphe,PrattLR:Quatst,lrp:apc02,Pratt:SCMF03}.
Then we draw upon  earlier calculations on H$^+$(aq) \cite{lrp:jpca02},
Li$^+$(aq) \cite{Pratt:ES:99,lrp:jacs00}, and Na$^+$(aq)
\cite{lrp:fpe01} and demonstrate that the predicted hydration state
provides a satisfactory description of the pair hydration free energies
for HOH, LiOH, and NaOH.   The involvement of HO$^-$(aq) in the
speciation of Be$^{2+}$(aq) has been studied recently by the same
methods  \cite{asthagir:cpl03}.

The H$^+$(aq) and HO$^-$(aq) ions are undoubtedly the most important
ions in aqueous phase chemistry, and particularly biological chemistry.
This is due largely to the fact that they are intrinsic to the aqueous
media and common extrinsic occupants of those phases  encounter these species.

The  Be$^{2+}$(aq) example noted above provides a specific motivation
for the present work.  Beryllium is an technologically important metal,
but inhaled beryllium dust is toxic and causes chronic beryllium disease
in a subset of exposed individuals. This disease eventually leads to
lung failure and is presently incurable. It is well-established that the
HO$^-$ ions can cross-link Be$^{2+}$ ions \cite{vacca:aic00}. These
colloidal beryllium-hydroxide clusters might ultimately trigger chronic
beryllium disease.

Understanding the hydration state of HO$^-$(aq) is also a key to
understanding its anomalously high diffusivity
\cite{Bernal:JCP33,Stillinger:tc78,asthagir:02}. Indeed both H$^+$ and
HO$^-$ are thought to diffuse by a proton/hole-hopping mechanism, and
this has implications that extend from biological transport of these
ions to the transport mechanisms in the polyelectrolyte membranes \cite
{EikerlingM:Defspt}  involved in some fuel cell designs.

\section{Quasi-chemical Theory}

In the quasi-chemical approach \cite{lrp:apc02}, the region around the
solute of interest is partitioned into inner and outer shell domains.
For the case of HO$^-$, the inner shell comprises the water molecules
directly coordinated with the ion. This domain can be treated treated
quantum mechanically, while the outer shell contributions can be
assessed using classical force-fields or dielectric continuum models.
The theory permits a variational check of the inner-outer partition
\cite{lrp:fpe01}, but this aspect has not been pursued here.  In the
present study outer shell contributions have been evaluated with a
dielectric continuum model and the trends confirmed by  molecular
dynamics calculations using classical interatomic potentials.

The inner shell reactions are:
\begin{eqnarray}
\mathrm{HO}^- + \; n\,\mathrm{H_2O} \rightleftharpoons \mathrm{HO[H_2O]}_n{}^-
\label{rxn}
\end{eqnarray}
The free energy change for these reactions were calculated using the
Gaussian programs \cite{gaussian}. The $\mathrm{HO\cdot[H_2O]_n{}^-}$ (n
= 0$\ldots$4) clusters were geometry optimized in the gas phase using
the B3LYP hybrid density functional\cite{b3lyp} and the 6-31+G(d,p)
basis set. Frequency calculations confirmed a true minimum, and the zero
point energies were computed at the same level of theory. Single point
energies were calculated with the 6-311+G(2d,p) and the aug-cc-pVTZ
basis sets, although most of our results pertain to the former basis.

For estimating the outer shell contribution, the ChelpG method
\cite{breneman:jcc90} was used to obtain partial atomic charges. Then
with the radii set developed by Stefanovich et
al.\cite{Stefanovich:cpl95}, surface tessera were generated
\cite{sanner}, and the hydration free energies of the clusters were
calculated using a dielectric continuum model \cite{lenhoff:jcc90}. With
this information and the binding free energies for the chemical
reactions, a primitive quasi-chemical approximation to the excess
chemical potential of HO$^-$(aq) in water is:
\begin{eqnarray}
\beta \mu_{\mathrm{OH}^-(aq)}^{ex} &\approx& - \ln \left\lbrack
\sum_{n\ge 0} \tilde{K}_n \rho_{\mathrm{H}_2\mathrm{O} }{}^n  \right
\rbrack \label{eq:regrouped}
\end{eqnarray}
where $\tilde{K}_n=K_n^{(0)}\exp\left[{-\beta
\left(\mu_{\mathrm{HO}(\mathrm{H}_2\mathrm{O})_n{}^-}^{ex}-n
\mu_{\mathrm{H}_2\mathrm{O} }^{ex}\right)}\right]$. $K_n^{(0)}$ is the
equilibrium constant for the reaction Eq.~\ref{rxn} in an ideal gas state, $n$ is the
hydration number of the most probable inner shell cluster, and
$\mathrm{\beta=1/k_\mathrm{B}T}$. The density factor
$\mathrm{\rho_{H_2O}}$ appearing in EQ.~\ref{eq:regrouped} reflects the
actual density of liquid water and its effect is included by a
replacement contribution of $-n k_\mathrm{B}T \ln
(\rho_\mathrm{H_2O}/\rho_0)$ = $-n k_\mathrm{B} T \ln (1354) $, where
$\rho_\mathrm{H_2O} = 1~$gm/cm$^3$ and $\rho_0 = 1\,\mathrm{atm}/RT$. (A
detailed discussion on standard states and this replacement contribution
can be found in Grabowski et al.~\cite{lrp:jpca02}.) Note 
EQ.~\ref{eq:regrouped} is a simplification of the broader theory, and
approximations enter at that stage.  But all these approximations are
available for scrutiny and improvement \cite{lrp:jpca02}.

\begin{figure}
\begin{center}
\includegraphics[width=5.25in]{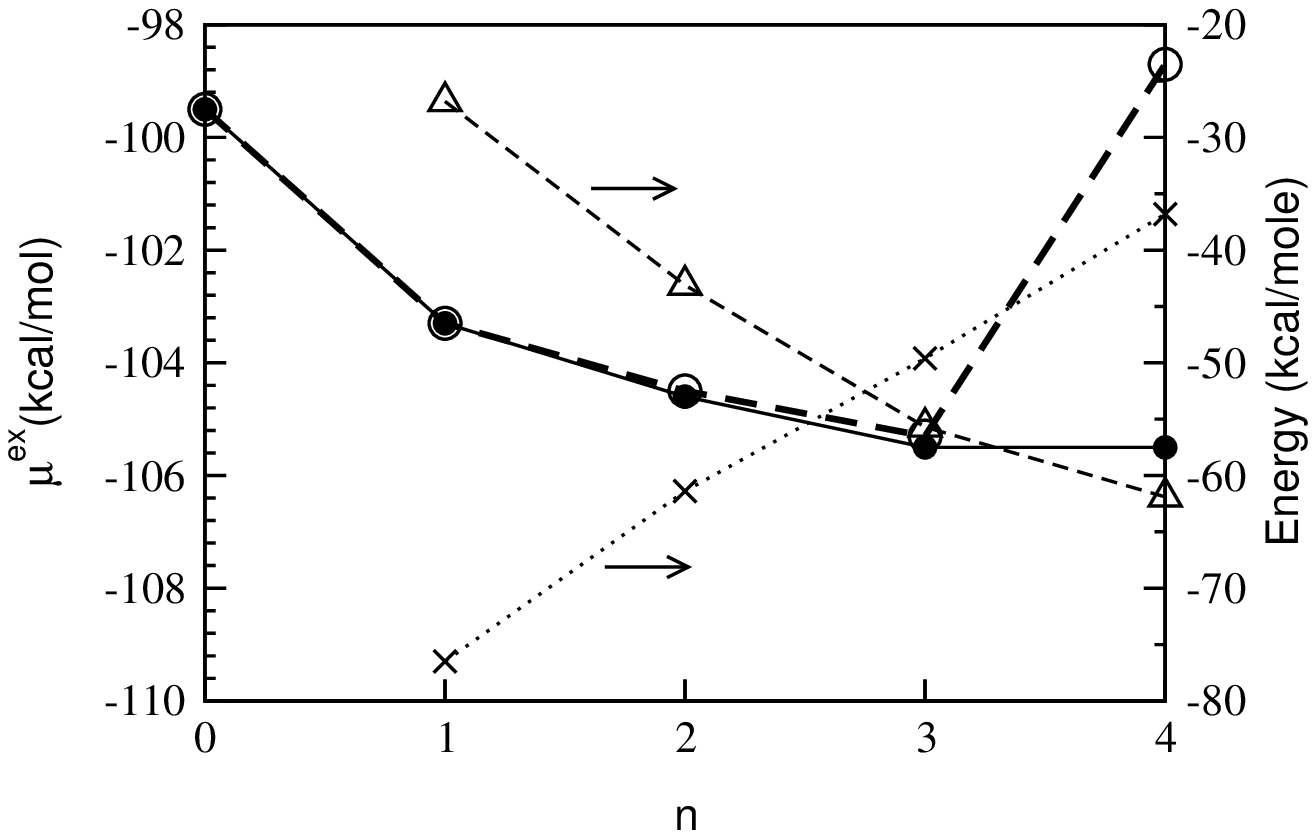}
\end{center}
\caption{Quasi-chemical contributions to the hydration free energy of
HO$^-$(aq) as a function of the inner-shell coordination number $n$.
$\circ$:  contribution of individual terms $\mathrm{-k_BT} \ln
\left\lbrack\tilde{K}_n \rho_{\mathrm{H}_2\mathrm{O} }{}^n\right\rbrack$
to $\mu^{ex}$;  see Ref.~\cite{lrp:jacs00}. $\bullet$: $-\mathrm{k_BT}
\ln \left\lbrack\sum_{m=0}^{m=n} \tilde{K}_m
\rho_{\mathrm{H}_2\mathrm{O} }{}^m  \right\rbrack$.  $\bigtriangleup$:
$-RT\ln K_n^{(0)} - n RT\ln \left\lbrack1354\right\rbrack$; $\times$:
$\mu_{\mathrm{HO}(\mathrm{H}_2\mathrm{O})_n{}^{-}}^{ex}-n
\mu_{\mathrm{H}_2\mathrm{O}}$. An observation volume of radius 1.7~{\AA}
centered on the anionic oxygen defined the inner shell. Change of that
radius, say to 2.0~{\AA}, would change the $n=0$ contribution roughly by
factor of (1.7/2.0).  But that wouldn't change the net result
substantially since the $n=3$ contribution dominates and the ion is
nearly buried by the ligands in that case.  This is an example of the
variational character of  the quasi-chemical theory noted
above.}\label{fg:hoqca} \end{figure}

In Table~\ref{tb:ener} the relevant energies are  collected, and
Fig.~\ref{fg:hoqca} gives the hydration free energy of the hydroxide
anion for various hydration states.  In order of decreasing stability
$\mathrm{HO\cdot[H_2O]_3{}^-} > \mathrm{HO\cdot[H_2O]_2{}^-} \sim
\mathrm{HO\cdot[H_2O]_1{}^-} > \mathrm{HO\cdot[H_2O]_4{}^-}$ is found.
The greater stability of $\mathrm{HO\cdot[H_2O]_3{}^-}$ is independent
of the level of theory;  calculations with the much larger aug-cc-pVTZ
basis give the same trends. Clearly (Fig.~\ref{fg:hoqca}) including the
$\mathrm{HO\cdot[H_2O]_4{}^-}$ contribution to the sum does not
appreciably alter the final excess chemical potential of HO$^-$(aq): the
whole effect is due accurately to the $\mathrm{HO\cdot[H_2O]_3{}^-}$
quasi-component.
\begin{table}
\caption{Electronic energy (a.u.), corrections (a.u.) to the
free energy for zero-point and thermal motions, and excess chemical potential (kcal/mole) using dielectric
continuum approximation with charges obtained at
B3LYP/6-311+G(2d,p). \bigskip}\label{tb:ener}
\begin{center}
\begin{tabular}{lrrr}\hline
                & \multicolumn{1}{c}{E} & \multicolumn{1}{c}{G$_{corr}$}  & $\mu^*$ \\ \hline
H$_2$O     &  $-76.45951$ &  $0.00298$  & $-7.7$  \\ 
HO$^-$      &  $-75.82779$ & $-0.00771$ & --- \\
$\mathrm{HO[H_2O]^{-}}$  & $-152.33413$  & 0.00634 & $-84.2$ \\
$\mathrm{HO[H_2O]_2{}^{-}}$ & $-228.83014$ & 0.02655 & $-76.8$ \\
$\mathrm{HO[H_2O]_3{}^{-}}$ & $-305.32036$ & 0.04705 & $-72.7$ \\
$\mathrm{HO[H_2O]_4{}^{-}}$ & $-381.80433$ & 0.07149 & $-67.6$ \\ 
$\mathrm{HO[H_2O]_3{}[H_2O]^{-}}$ & -381.80104 & 0.06450 & --- \\ \hline
\end{tabular}
\end{center}
\end{table}

A comparative  rationalization of the electronic structure results on
$\mathrm{HO\cdot[H_2O]_3{}^-}$ and $\mathrm{HO\cdot[H_2O]_4{}^-}$ is the
following:  The nominal hydroxide hydrogen atom in these {\em negative}
ions is less positively charged than is typical of  water hydrogens.  As
a result, opportunities for hydrogen bond donation to that nominal
hydroxide hydrogen have diminished profitability.  The fourth water
ligand then prefers to crowd among the other three on the oxygen side of
the hydroxide anion.

To assess the  influence of the level of theory used, we have calculated
the free energy change for the reaction:
\begin{eqnarray}
\mathrm{HO\cdot[H_2O]_3{}^- + H_2O \rightleftharpoons
HO\cdot[H_2O]_4{}^-} \label{eq:inneradd}
\end{eqnarray}
using the BLYP, PW91, and PBE density functionals, with geometry
optimizations at those levels.  MP2 single point calculations have been
performed on the B3LYP derived geometries as well.  An important feature
that emerges from Fig.~\ref{fg:hoqca} and Table~\ref{tb:comps}  is that outer shell contribution
favors $\mathrm{HO\cdot[H_2O]_3{}^-}$, at all levels of theory, principally by providing an
assessment of the hydration of the separated ligands.

\begin{table}
\caption{Contributions to the free energy change of
reaction~\ref{eq:inneradd}. $\mathrm{\Delta G^\circ}$ is the free energy
change in the standard  1~atm pressure ideal gas state. $\mathrm{\Delta
G}$ accounts for the concentration of water (55.5~M), and the
concentration of the quasi-components (1~M ideally diluted).
$\Delta\mu^{ex}$ is the change in the excess chemical potential.
$\mathrm{\Delta G(aq)}$ is the net free energy change. \bigskip}\label{tb:comps}
\begin{center}
\begin{tabular}{lrrrr}\hline
Theory &  $\Delta G^\circ$ & $\Delta G$ & $\Delta\mu^{ex}$ &  $\mathrm{\Delta G(aq)}$ \\ \hline
B3LYP &  -1.9  & -6.2 & 12.8  & 6.6 \\
BLYP   &  -2.0  &  -6.3 & 10.3 & 4.0  \\
PW91  &  -1.6  &  -5.9 &  13.5 & 7.6 \\
PBE     &  -1.1  &  -5.4 &  13.4 &  8.0 \\  
MP2    &   -4.6  &  -8.9 & 13.6 &  4.7  \\ \hline
\end{tabular}
\end{center}
\end{table}

To check limitations of the dielectric continuum model for outer shell
contributions, the charging free energies of
$\mathrm{HO\cdot[H_2O]_3{}^-}$ and $\mathrm{HO\cdot[H_2O]_4{}^-}$ were
obtained using classical molecular dynamics with TIP3P  potential for
water and the TIP3P van~der~Waals parameters for the oxygen and hydrogen
atoms of the quasi-component.  We find that
$\mu^{ex}_{\mathrm{HO\cdot[H_2O]_3{}^-}}-
\mu^{ex}_{\mathrm{HO\cdot[H_2O]_4{}^-}} = -6.9$~kcal/mol in reasonable
agreement with the $-6$~kcal/mol found using a dielectric model.
Importantly, the trend is unaltered. (See \cite{lrpions:jcp03} for
details on the classical simulation procedure.) Positive outer shell
packing contributions are not addressed here, but these are expected to
be slightly larger for $\mathrm{HO\cdot[H_2O]_4{}^-}$ than for
$\mathrm{HO\cdot[H_2O]_3{}^-}$ and hence should enhance the calculated
difference.

\section{Discussions}

Fig.~\ref{fg:hoqca}  and Table~\ref{tb:comps} show that
contributions for inner shell water additions to
$\mathrm{HO\cdot[H_2O]_3{}^-}$ are in fact favorable;  this is routinely
observed \cite{lrp:jacs00} and  implies that ligand hydration typically
plays a signficant role in establishing the probable coordination
numbers.

Another common observation in applying a quasi-chemical approach to
ion hydration problems is that aggregates beyond the most probable size
begin to find favorable outer shell placements for the later additions.
This seems to be the case in the present problem too. Alternative
arrangements of four water molecules, such as
$\mathrm{HO\cdot[H_2O]_3\cdot[H_2O]{}^-}$, are  more favorable than
$\mathrm{HO\cdot[H_2O]_4{}^-}$ by about 2~kcal/mole
(Table~\ref{tb:speller} ). Numerous such arrangments are possible
\cite{chaudhuri:mp01}.  In the specific case we have considered, the
fourth water molecule hydrogen-bonds with the inner shell water
molecules, similar to structure OHW4III in Fig.~2 of
\cite{chaudhuri:mp01}.

Table~\ref{tb:speller}  shows that the present inner shell computations
are in reasonable agreement with experimental results. They are also in
reasonable agreement with theoretical calculations reported in
\cite{chaudhuri:mp01} with the differences attributable to differences
in the basis sets for minimization and energy evaluations.  Note that
within the primitive quasi-chemical approach,  outer-shell structures of
the fourth water  are excluded here because they are accounted for in outer
shell contributions. Those arrangements are in fact part of the
outer-sphere arrangements of the $\mathrm{HO\cdot[H_2O]_3{}^-}$
quasi-component. For the n=4 case, one expects many isomers to be
present, but thermochemical measurements obviously cannot  specify the
structure. The trends (Table~\ref{tb:speller}) strongly suggest that in
the gas-phase the n=4 cluster in fact must involve an outer-sphere
arrangement of the fourth water. This shell-closure at n=3 was inferred
by Moet-Ner and Speller \cite{speller:jpc86} based on their
thermochemical analysis of the step-wise attachment of water to HO$^-$.
These authors also noted possible artifacts in a much earlier work
\cite{kebarle:jpc70} that did not show the shell effect.
\begin{table}[h]
\caption{Standard free energy ($\mathrm{\Delta G^\circ}$) for adding $n$
water molecules to HO$^-$.  The experimental results (with an error bar
of about $\pm$~1.5~kcal/mole) are from \cite{speller:jpc86}.  A: This
work at B3LYP/6-311+G(2d,p). B: This work at B3LYP/aug-cc-pVTZ. The case
$3+1$ refers to the outer-sphere arrangement of the fourth water
discussed in the text. Theoretical calculations for comparable
structures from \cite{chaudhuri:mp01} at the B3LYP/aug-cc-pVDZ level are
also shown.\bigskip}\label{tb:speller}
\begin{center}
\begin{tabular}{crrrr}\hline
$n$ & Expt & A & B  & \cite{chaudhuri:mp01}  \\ \hline
 1    & -20.0 & -22.5 & -20.5 & -20.5 \\
 2   & -31.2 & -34.5 &   -31.1   & -32.2 \\
 3  & -40.2 & -42.8  & -38.0 & -40.1 \\
 3+1 &   &  -47.0 &   -41.4        &  -43.2 \\ 
 4  &  \raisebox{1.5ex}[0pt]{(-46.0)} & -44.7 &  -38.3 & -39.0  \\ \hline
\end{tabular}
\end{center}
\end{table}

The lower energy of the outer shell arrangement of the fourth water was
recently confirmed  spectroscopically by Johnson and coworkers
\cite{johnson:sc03}.   Those experiments showed that  shell closure by
the ligating water molecules occurs when three water molecules are
hydrogen bonded to the HO$^-$ ion.  New spectral features appeared with the addition of a fourth water molecule.  These new features were a result of hydrogen bonding of the fourth water molecule to first solvation shell waters instead of direct coordination with HO$^-$.

The conclusion that  $\mathrm{HO\cdot[H_2O]_3{}^-}$ is the  predominant
inner shell  hydration number in liquid water under standard conditions
has been supported by \emph{ab initio} molecular dynamics (AIMD) 
simulations \cite{asthagir:02,scandolo:jcp02,voth:vIIjcp01}. But
thermodynamic tests can be considered too. Earlier quasi-chemical
studies have confirmed the dominant coordination structure of H$^+$,
Li$^+$ and Na$^+$ \cite{lrpions:jcp03,lrp:jpca02}. The dominant
coordination structure of Li$^+$ and Na$^+$ were also cross-checked
against Born-Oppenheimer AIMD \cite{lrp:jacs00,lrp:fpe01}. Thus we can
compute the hydration free energy for the neutral ion combinations HOH,
LiOH, and  NaOH.  These pair hydration free energies, obtained using
classical molecular dynamics simulations for the outer shell
contributions, are given in Table~\ref{tb:pair} (after Table~III in
\cite{lrpions:jcp03}). The dominant $\mathrm{HO\cdot[H_2O]_3{}^-}$
structure was also found to be the best descriptor for the ionization of
water \cite{lrpions:jcp03} and of $\mathrm{Be[H_2{}O]_4{}^{2+}}$
\cite{asthagir:cpl03}.

\begin{table}
\caption{Solvation free energy of neutral ion pairs (kcal/mole). The
solutes are transferred from 1~M (ideal gas) to 1~M (ideally diluted
solute). TIP3 and SPC/E  refer to the potentials used for the water
model. The pair hydration free energy for HOH was obtained based on the
experimental gas phase free energy of dissociation (383.7~kcal/mole)
\cite{bartmess}, the known pK of water (15.7) \cite{pearson:jacs86} and
the hydration free energy of water (-6.3~kcal/mole)  obtained from
phase-equilibria information. The values for LiOH and NaOH are from
\cite{coe:jpca98}.\bigskip }\label{tb:pair}
\begin{center}
\begin{tabular}{lrrr} \hline
  &  TIP3P & SPC/E & Expt \\ \hline
HOH & -366.4 & -367.8 & -368.1 \\
LiOH  & -236.2 & -236.5 & -233.3 \\
NaOH & -211.1 & -212.5 & -208.1 \\ \hline
\end{tabular}
\end{center}
\end{table}

The agreement shown in Table~\ref{tb:pair} is excellent. The minor
discrepancies are attributable to neglect of a variety of secondary
effects:  anharmonicity of the quasi-component structures, packing
effects, and dispersion interactions. An assumption that 
$\mathrm{HO\cdot[H_2O]_4{}^-}$ was the dominant form would have
predicted these hydration free energies to be significantly more
positive than the experimental results.

\section{Conclusions}
The  present quasi-chemical theory applied to HO$^-$(aq) leads to the
conclusion  that $\mathrm{HO\cdot[H_2O]_3{}^-}$ is the dominant inner
shell coordination structure for  the HO$^-$(aq) ion in liquid water
under standard conditions. The $\mathrm{HO\cdot[H_2O]_4{}^-}$ is less
favorable by nearly 7~kcal/mole ($\approx 12~\mathrm{k_BT}$). The
prediction of $\mathrm{HO\cdot[H_2O]_3{}^-}$  as the dominant form has
been successfully used in predicting the hydration free energies of
neutral ion combinations HOH, LiOH, and  NaOH. Thus based on different
lines of investigation, we conclude that HO$^-$ is predominantly
$\mathrm{HO\cdot[H_2O]_3{}^-}$ in liquid water.

\section*{Acknowledgements}
The work at Los Alamos was supported by the US Department of Energy,
contract W-7405-ENG-36, under the LDRD program at Los Alamos.
LA-UR-03-3473.


\end{document}